\documentclass[10pt,a4paper]{article}
\usepackage{amsmath}
\usepackage{amssymb}
\usepackage{amsfonts}
\usepackage{amstext}
\usepackage{amsbsy}
\usepackage[mathscr]{eucal}
\usepackage{graphicx}
\usepackage{color}
\newcommand{\beq}{\begin{equation}}
\newcommand{\eeq}{\end{equation}}
\newcommand{\beqa}{\begin{eqnarray}}
\newcommand{\eeqa}{\end{eqnarray}}
\newcommand{\ket}[1]{| #1 \rangle}
\newcommand{\bra}[1]{\langle #1 |}

\title{\Large\textbf{A toric varieties approach to geometrical structure of multipartite states}}

\author{\textit{ Hoshang Heydari}\\
        \small\textit{Physics Department, Stockholm university 10691 Stockholm Sweden}\\
\\\small\textit{Email: hoshang@fysik.su.se}}
\begin{document}
\maketitle

\begin{abstract}
 We investigate the geometrical structures of multipartite states based on construction of  toric varieties.
In particular, we describe pure quantum systems in terms of affine toric varieties and projective embedding of these varieties in complex projective spaces. We show that a quantum system can be corresponds to a toric variety of a fan which is constructed by gluing together affine toric varieties of polytopes. Moreover, we show that the projective toric varieties are the spaces of separable multipartite quantum states. The construction
 is a generalization of the complex multi-projective Segre variety. Our construction suggests a systematic way of looking at the structures of multipartite quantum systems.
\end{abstract}


\maketitle

\section{Introduction}
\label{intro}
Recently, the geometrical structures of multipartite quantum systems have became an active part
of ongoing research on foundations of quantum mechanics with applications in the field of quantum information and
quantum computation. On the other hand toric varieties are important structures in algebraic geometry \cite{Ewald,GKZ,Fulton,Proud05} which have also some applications  in the field of string theory \cite{Vafa,Cox}. In this paper, we will  establish a relation between multipartite systems and toric  varieties. In particular,  we review the construction of affine and projective toric varieties.
   Then, we discuss the geometrical structures of multipartite systems in terms of toric varieties. We also assume that  the readers are familiar with  basic definitions and properties of abstract varieties in algebraic geometry \cite{Griff78,Hart77,Mum76}.
    In this paper, we will denote a general, multipartite quantum system with $m$
subsystems by
$\mathcal{Q}$ $
=\mathcal{Q}_{1}\mathcal{Q}_{2}\cdots\mathcal{Q}_{m}$, consisting
of a state
\begin{equation}\label{Mstate}
\ket{\Psi}=\sum^{N_{1}-1}_{k_{1}=0}\sum^{N_{2}-1}_{k_{2}=0}\cdots\sum^{N_{m}-1}_{k_{m}=0}
\alpha_{k_{1}k_{2}\cdots k_{m}} \ket{k_{1}k_{2}\cdots k_{m}}
\end{equation}
 and a density operator
$\rho_{\mathcal{Q}}=\sum^{\mathrm{N}}_{i=1}p_{i}\ket{\Psi_{i}}\bra{\Psi_{i}}$,
for all $0\leq p_{i}\leq 1$ and $\sum^{\mathrm{N}}_{i=1}p_{i}=1$,
 acting on the Hilbert space $
\mathcal{H}_{\mathcal{Q}}=\mathcal{H}_{\mathcal{Q}_{1}}\otimes
\mathcal{H}_{\mathcal{Q}_{2}}\otimes\cdots\otimes\mathcal{H}_{\mathcal{Q}_{m}},
$ where the dimension of the $j$th Hilbert space is given  by
$N_{j}=\dim(\mathcal{H}_{\mathcal{Q}_{j}})$.

\section{Toric varieties}
\label{sec:2}
The construction of toric varieties usually are based on two different branches of mathematics, namely, combinatorial geometry and algebraic geometry. Here, we will review the   basic notations and structures of toric varieties.
Let $\mathcal{C}$ be a complex algebraic field. Then, an affine
$n$-space over $\mathcal{C}$ denoted $\mathcal{C}^{n}$ is the set of
all $n$-tuples of elements of $\mathcal{C}$. An element
$P\in\mathcal{C}^{n}$ is called a point of $\mathcal{C}^{n}$ and if
$P=(a_{1},a_{2},\ldots,a_{n})$ with $a_{j}\in\mathcal{C}$, then
$a_{j}$ is called the coordinates of $P$.
 A complex projective space $\mathcal{P}_{\mathcal{C}}^{n}$ is
defined to be the set of lines through the origin in
$\mathcal{C}^{n+1}$, that is, $
\mathcal{P}_{\mathcal{C}}^{n}=\frac{\mathcal{C}^{n+1}-\{0\}}{
(x_{0},\ldots,x_{n})\sim(y_{0},\ldots,y_{n})},~\lambda\in
\mathcal{C}-0,~y_{i}=\lambda x_{i}$ for all $0\leq i\leq n $.

A general toric variety is an irreducible variety $\mathcal{X}$ that satisfies the following conditions. First of all $(\mathcal{C}^{*})^{n}$ is a Zariski open subset of $\mathcal{X}$ and the action of
 $(\mathcal{C}^{*})^{n}$  on itself can extend to an action of  $(\mathcal{C}^{*})^{n}$  on the variety
 $\mathcal{X}$. As an example we will show that the complex projective space $\mathcal{P}^{n}$ is a toric variety. If $z_{0},z_{1},
 \ldots,z_{n}$ are homogeneous coordinate of  $\mathcal{P}^{n}$. Then, the map
 $(\mathcal{C}^{*})^{n}\longrightarrow\mathcal{P}^{n}$ is defined by $(t_{1},t_{2},
 \ldots,t_{n})\mapsto(1,t_{1},
 \ldots,t_{n})$  and we have
\begin{equation}
 (t_{1},t_{2},
 \ldots,t_{n})\cdot (a_{0},a_{1},
 \ldots,a_{n})=(a_{0},t_{1}a_{1},
 \ldots,t_{n}a_{n})
\end{equation} which proof our claim that $\mathcal{P}^{n}$ is a toric variety.
 We can also define toric varieties with combinatorial information such as polytope and fan.
 But first we will give a short introduction to the basic of combinatorial geometry which is important in definition of toric varieties. Let $S\subset \mathcal{R}^{n}$ be finite subset, then a convex polyhedral cone is defined by
$
 \sigma=\mathrm{Cone}(S)=\left\{\sum_{v\in S}\lambda_{v}v|\lambda_{v}\geq0\right\}.
$
In this case $\sigma$ is generated by $S$.  In a similar way we define  a polytope by
$
 P=\mathrm{Conv}(S)=\left\{\sum_{v\in S}\lambda_{v}v|\lambda_{v}\geq0, \sum_{v\in S}\lambda_{v}=1\right\}.
$
We also could say that $P$ is convex hull of $S$. A convex polyhedral cone is called simplicial if it is generated by linearly independent set. Now, let $\sigma\subset \mathcal{R}^{n}$ be a convex polyhedarl cone and $\langle u,v\rangle$ be a natural pairing between $u\in \mathcal{R}^{n}$ and $v\in\mathcal{R}^{n}$. Then, the dual cone of the $\sigma$ is define by
$
 \sigma^{\wedge}=\left\{u\in \mathcal{R}^{n*}|\langle u,v\rangle\geq0~\forall~v\in\sigma\right\},
$
where $\mathcal{R}^{n*}$ is dual of $\mathcal{R}^{n}$. We also define the polar of $\sigma$ as
 $
 \sigma^{\circ}=\left\{u\in \mathcal{R}^{n*}|\langle u,v\rangle\geq-1~\forall~v\in\sigma\right\}
$. We call a convex polyhedral cone strongly convex if $\sigma\cap(-\sigma)=\{0\}$.

Next we will define rational polyhedral cones. A free Abelian group of finite rank is called a lattice, e.g., $N\simeq\mathcal{Z}^{n}$. The dual of a lattice $N$ is defined by
$M=\mathrm{Hom}_{\mathcal{Z}}(N,\mathcal{Z})$ which has rank $n$. We also define a vector space and its dual by $N_{\mathcal{R}}=N\otimes_{\mathcal{Z}}\mathcal{R}\simeq \mathcal{R}^{n}$ and $M_{\mathcal{R}}=M\otimes_{\mathcal{Z}}\mathcal{R}\simeq \mathcal{R}^{n*}$ respectively.
 Moreover, if $\sigma=\mathrm{Cone}(S)$ for some finite set $S\subset N$, then $\sigma\subset N_{\mathcal{R}}$ is a rational polydehral cone. Furthermore, if  $\sigma\subset N_{\mathcal{R}}$ is a rational polyhedral cone, then   $S_{\sigma}=\sigma^{\wedge}\cap M$ is a semigroup under addition with $0\in S_{\sigma}$ as additive identity which is finitely generated by Gordan's lemma.

Here we will define a fan which is important in the construction of toric varieties. Let $\Sigma\subset N_{\mathcal{R}}$ be a finite non-empty set of strongly convex rational polyhedral cones. Then $\Sigma$ is called a fan if each face of a cone in $\Sigma$  belongs to $\Sigma$ and the intersection of any two cones in $\Sigma$  is a face of each.

 Now, we can obtain the coordinate ring of a variety by associating to the semigroup $S$ a finitely generated commutative $\mathcal{C}$-algebra without nilpotent as follows. We associate  to an arbitrary additive semigroup its semigroup algebra $\mathcal{C}[S]$ which as a vector space has the set $S$ as basis. The elements of $\mathcal{C}[S]$ are linear combinations
 $\sum_{u\in S}a_{u}\chi^{u}$ and the product in $\mathcal{C}[S]$ is determined by the addition in $S$ using  $\chi^{u}\chi^{u^{'}}=\chi^{u+u^{'}}$ which is called the exponential rule. Moreover, a set of semigroup generators $\{u_{i}: i\in I\}$ for $S$ gives algebra generators $\{\chi^{u_{i}}: i\in I\}$ for  $\mathcal{C}[S]$.

 Now, let $\sigma\subset N_{\mathcal{R}}$ be a strongly convex rational polyhedral cone and  $A_{\sigma}=\mathcal{C}[S_{\sigma}]$ be an algebra which is a normal domain. Then,
\begin{equation}
\mathcal{X}_{\sigma}=\mathrm{Spec}(\mathcal{C}[S_{\sigma}])=\mathrm{Spec}(A_{\sigma})
\end{equation}
is called a affine toric variety.   Next we need to define Laurent polynomials and monomial algebras. But first we observe that the dual cone $\sigma^{\vee}$ of the zero cone $\{0\}\subset N_{\mathcal{R}}$ is all of $ M_{\mathcal{R}}$ and the associated semigroup $S_{\sigma}$ is the group $M\simeq \mathcal{Z}^{n}$. Moreover, let $(e_{1},e_{2},\ldots,e_{n})$ be a basis of $N$ and
$(e^{*}_{1},e^{*}_{2},\ldots,e^{*}_{n})$ be its dual basis for $M$. Then, the elements $\pm e^{*}_{1},\pm e^{*}_{2},\ldots,\pm e^{*}_{n}$ generate $M$ as semigroup. The algebra of Laurent polynomials is defined by
\begin{equation}
\mathcal{C}[z,z^{-1}]=\mathcal{C}[z_{1},z^{-1}_{1},\ldots,z_{n},z^{-1}_{n}],
\end{equation}
where $z_{i}=\chi^{e^{*}_{i}}$. The terms  of the form $\lambda \cdot z^{\beta}=\lambda z^{\beta_{1}}_{1}z^{\beta_{2}}_{2}\cdots z^{\beta_{n}}_{n}$ for $\beta=(\beta_{1},\beta_{2},\ldots,\beta_{n})\in \mathcal{Z}$ and $\lambda\in \mathcal{C}^{\times}$ are called Laurent monomials. A ring $R$ of Laurent polynomials is called a monomial algebra if it is a $\mathcal{C}$-algebra generated bye Laurent monomials. Moreover, for a lattice cone $\sigma$, the ring
$R_{\sigma}=\{f\in \mathcal{C}[z,z^{-1}]:\mathrm{supp}(f)\subset \sigma\}
$
is a finitely generated monomial algebra, where the support of a Laurent polynomial $f=\sum\lambda_{i}z^{i}$ is defined by
$\mathrm{supp}(f)=\{i\in \mathcal{Z}^{n}:\lambda_{i}\neq0\}.
$ Now, for a lattice cone $\sigma$ we can define an affine toric variety to be the maximal spectrum $\mathcal{X}_{\sigma}=\mathrm{Spec}R_{\sigma}$.  A toric variety
$\mathcal{X}_{\Sigma}$ associated to a fan $\Sigma$ is the result of gluing affine varieties
$\mathcal{X}_{\sigma}=\mathrm{Spec}R_{\sigma}$ for all $\sigma\in \Sigma$ by identifying $\mathcal{X}_{\sigma}$ with the corresponding Zariski open subset in $\mathcal{X}_{\sigma^{'}}$ if
$\sigma$ is a face of $\sigma^{'}$. That is,
first we take the disjoint union of all affine toric varieties $\mathcal{X}_{\sigma}$ corresponding to the cones of $\Sigma$. Then by gluing all these affine toric varieties together we get $\mathcal{X}_{\Sigma}$.


Next, we will give a short introduction to an embedding of a toric variety into a projective space $\mathcal{P}^{r}$.
A compact toric variety $\mathcal{X}_{\Sigma}$ is called projective if there exists an injective morphism
$$\Phi:\mathcal{X}_{\Sigma}\longrightarrow\mathcal{P}^{r}$$ of $\mathcal{X}_{\Sigma}$
 into some projective space such that $\Phi(\mathcal{X}_{\Sigma})$ is Zariski
 closed in $\mathcal{P}^{r}$.
A toric variety $\mathcal{X}_{\Sigma}$ is equivariantly projective if and only if $\Sigma$ is strongly polytopal. Now, let $\mathcal{X}_{\Sigma}$ be equivariantly projective and morphism
$\Phi$  be embedding which is induced by the rational map $\phi:\mathcal{X}_{\Sigma}  \longrightarrow  \mathcal{P}^{r}$
defined by $p \mapsto[z^{m_{0}},z^{m_{1}},\ldots,z^{m_{r}}],$ where $z^{m_{l}}(p)=p^{m_{l}}$ in case $p=(p_{1},p_{2},\ldots p_{n})$. Then, the rational map $\Phi(\mathcal{X}_{\Sigma})$ is the set of common solutions of finitely many monomial equations
$
x^{\beta_{0}}_{i_{0}}x^{\beta_{1}}_{i_{1}}\cdots x^{\beta_{k}}_{i_{k}}=x^{\beta_{k+1}}_{i_{k+1}}x^{\beta_{k+2}}_{i_{k+2}}\cdots x^{\beta_{r}}_{i_{r}}
$
which satisfy the following relationships
$
  \beta_{0}m_{0}+\beta_{1}m_{1}+\cdots +\beta_{k}m_{k}=\beta_{k+1}m_{k+1}+\beta_{k+2}m_{k+2}+\cdots +\beta_{r}m_{r}$ and $
  \beta_{0}+\beta_{1}+\cdots +\beta_{k}=\beta_{k+1}+\beta_{k+2}+\cdots +\beta_{r}
$,
for all $\beta_{l}\in \mathcal{Z}_{\geq 0}$ and $l=0,2,\ldots, r$ \cite{Ewald}. This construction of projective toric variety is very important in the next section.


\section{Multipartite systems and toric varieties}
\label{sec:3}
After going through a lot of preparation which includes many definition and abstraction from pure mathematics we are ready to discuss the geometrical structures of multipartite quantum systems based on toric varieties.
For multi-qubit state $\ket{\Psi}=\sum^{1}_{k_{1},\ldots,k_{m}=0}
\alpha_{k_{1}\cdots k_{m}} \ket{k_{1}\cdots k_{m}}$ the space of the separable state is given by the Segre embedding of
\begin{eqnarray}
\mathcal{CP}^{1}\times\mathcal{CP}^{1}\times\cdots\times\mathcal{CP}^{1}&=&
\{((\alpha^{1}_{0},\alpha^{1}_{1}),\ldots,(\alpha^{m}_{0},\alpha^{m}_{1}))):\\\nonumber&& (\alpha^{1}_{0},\alpha^{1}_{1})\neq(0,0),~\ldots
,~(\alpha^{m}_{0},\alpha^{m}_{1})\neq(0,0)\}.
\end{eqnarray}
Let $M=\mathcal{Z}^{m}$ and consider the $m$ cube $P\subset M_{R}$ centered at the origin with vertices $(\pm1,\ldots,\pm1)$. This gives the toric variety $\mathcal{X}_{P}=
\mathcal{CP}^{1}\times\mathcal{CP}^{1}\times\cdots\times\mathcal{CP}^{1}$ and the polar $P^{\circ}\subset N_{R}$ is a polytope with vertices $\pm e_1,\ldots, \pm e_m$.  Now, let $z_{1}=\alpha^{1}_{1}/\alpha^{1}_{0},z_{2}=\alpha^{2}_{1}/\alpha^{2}_{0},\ldots z_{m}=\alpha^{m}_{1}/\alpha^{m}_{0}$. Then, we have the following map
$\Phi:\mathcal{X}_{P}^{m}\longrightarrow\mathcal{CP}^{2^{m-1}}$ which is defined  by
$$
(z_{1},z_{2},\ldots, z_{m})\longmapsto Z=(1,z_{1},z_{2},\ldots,z_{m},z_{1}z_{2},\ldots,z_{1}z_{2}\cdots z_{m}).
$$
Now, the map $\Phi(\mathcal{X}_{P})$ is  a set of the common
solutions of the following monomial equations
\begin{equation}
x^{\beta_{0}}_{i_{0}}x^{\beta_{1}}_{i_{1}}\cdots
x^{\beta_{2^{m-1}-1}}_{i_{2^{m-1}-1}}=x^{\beta_{2^{m-1}}}_{i_{2^{m-1}}}
\cdots x^{\beta_{2^{m}-1}}_{i_{2^{m}-1}}
\end{equation}
that gives  quadratic polynomials $\alpha_{k_{1}k_{2}\ldots
k_{m}}\alpha_{l_{1}l_{2}\ldots l_{m}} = \alpha_{k_{1}k_{2}\ldots
l_{j}\ldots k_{m}}\alpha_{l_{1}l_{2} \ldots k_{j}\ldots l_{m}}$ for
all $j=1,2,\ldots,m$ which  coincides with the Segre ideals.
Moreover, we have
\begin{equation}
\Phi(\mathcal{X}_{P})=\mathrm{Specm} \mathcal{C}[\alpha_{00\ldots
0},\alpha_{00\ldots 1},\ldots,\alpha_{11\ldots
1}]/\mathcal{I}(\mathcal{A}) ,
\end{equation}
where $\mathcal{I}(\mathcal{A})=\langle \alpha_{k_{1}k_{2}\ldots
k_{m}}\alpha_{l_{1}l_{2}\ldots l_{m}} - \alpha_{k_{1}k_{2}\ldots
l_{j}\ldots k_{m}}\alpha_{l_{1}l_{2} \ldots
k_{j}\ldots l_{m}}\rangle_{\forall j;k_{j},l_{j}=0,1}$.
This toric variety describes the space of separable states in a multi-qubit quantum systems.

\begin{thebibliography}{22}
\bibitem{Ewald}G. Ewald, {\it Combinatorial Convexity and Algebraic Geometry}, Springer.
\bibitem{GKZ} Gelfand, Kapranov, Zelevinsky, {\it Discriminants, resultants, and multidimensional determinants}, (1994).
\bibitem{Fulton} Fulton - {\it Introduction to Toric Varieties}, (1991)
\bibitem{Proud05}N. Proudfoot, eprint math.AG/0502366.
\bibitem{Vafa} Zaslow, eds, {\it Mirror Symmetry}, AMS, (2003).
\bibitem{Cox} D. Cox and S.  Katz, {\it Mirror symmetry and algebraic geometry}, AMS, (1999).

\bibitem{Griff78} P. Griffiths and J. Harris, {\it Principles of
  algebraic geometry}, Wiley and Sons, New York, 1978.
 \bibitem{Hart77} R. Hartshorne, {\it Algebraic Geometry}, Springer-Verlag, New York, 1977.
\bibitem{Mum76} D. Mumford, {\it Algebraic Geometry I,
Complex Projective Varieties}, Springer-Verlag, Berlin, 1976.
\end{thebibliography}

As an illustrative example we will in detail discuss a three-qubit state $\ket{\Psi}=\sum^{1}_{k_{1},k_{2},k_{3}=0}
\alpha_{k_{1}k_{2}k_{3}} \ket{k_{1}k_{2}k_{3}}.$ For this  state the separable state is given by the Segre embedding of
\begin{eqnarray}
\mathcal{CP}^{1}\times\mathcal{CP}^{1}\times\mathcal{CP}^{1}&=&
\{((\alpha^{1}_{0},\alpha^{1}_{1}),(\alpha^{2}_{0},\alpha^{2}_{1}),(\alpha^{3}_{0},\alpha^{3}_{1}))):\\\nonumber&&
 (\alpha^{1}_{0},\alpha^{1}_{1})\neq(0,0),~(\alpha^{2}_{0},\alpha^{2}_{1})\neq(0,0)
,~(\alpha^{3}_{0},\alpha^{3}_{1})\neq(0,0)\}.\end{eqnarray}
Moreover, let $M=\mathcal{Z}^{3}$ and consider the polytope $P$ centered at the origin with vertices $(\pm1,\pm1,\pm1)$ which is a cube. This gives the toric variety $\mathcal{X}_{P}=\mathcal{CP}^{1}\times\mathcal{CP}^{1}\times\mathcal{CP}^{1}$. We can get the fan of $\mathcal{X}_{P}$, by observing that the polar $P^{\circ}$ is the octahedron with vertices $\pm e_1,\pm e_2, \pm e_3$.  Now, for example, let $z_{1}=\alpha^{1}_{1}/\alpha^{1}_{0}$,
 $z_{2}=\alpha^{2}_{1}/\alpha^{2}_{0}$, and $z_{3}=\alpha^{3}_{1}/\alpha^{3}_{0}$. Then, the map
$\Phi:\mathcal{X}_{P}\longrightarrow\mathcal{CP}^{7}$ is defined by
\begin{equation}
(z_{1},z_{2},z_{3})\longmapsto(1,z_{1},z_{2},z_{3},z_{1}z_{2},z_{1}z_{3},z_{2}z_{3},z_{1}z_{2}z_{3}),
\end{equation}
where $(1,z_{1},z_{2},z_{3},z_{1}z_{2},z_{1}z_{3},z_{2}z_{3},z_{1}z_{2}z_{3})=
(\alpha^{1}_{0}\alpha^{2}_{0}\alpha^{3}_{0},\alpha^{1}_{1}\alpha^{2}_{0}\alpha^{3}_{0},
\ldots,\alpha^{1}_{1}\alpha^{2}_{1}\alpha^{3}_{1})$.
$\Phi(\mathcal{X}_{P})$ is the set of common solutions of
finitely many monomial equations
\begin{equation}
x^{\beta_{0}}_{i_{0}}x^{\beta_{1}}_{i_{1}}x^{\beta_{2}}_{i_{2}}x^{\beta_{3}}_{i_{3}}
=x^{\beta_{4}}_{i_{4}}x^{\beta_{5}}_{i_{5}}x^{\beta_{6}}_{i_{6}}x^{\beta_{7}}_{i_{7}}
\Longrightarrow\alpha_{k_{1}k_{2}k_{3}}\alpha_{l_{1}l_{2}l_{3}}
=\alpha_{k_{1}l_{j}k_{3}}\alpha_{l_{1}k_{j}l_{3}},
\end{equation}
where e.g., $\beta_{0}=\beta_{1}=\beta_{4}=\beta_{5}=1$,
$\beta_{2}=\beta_{3}=\beta_{6}=\beta_{7}=0$, and $j=1,2,3$. Then,
the projective toric variety is gives                                                                                                                                            by
\begin{equation}
\Phi(\mathcal{X}_{P})=\mathrm{Specm}
\mathcal{C}[\alpha_{000},\alpha_{001},\ldots,\alpha_{111}]/
\mathcal{I}(\mathcal{A}),
\end{equation}
where $\mathcal{I}(\mathcal{A})=\langle \alpha_{k_{1}k_{2}k_{3}}\alpha_{l_{1}l_{2}l_{3}} - \alpha_{k_{1}
l_{j} k_{3}}\alpha_{l_{1}
k_{j}l_{3}}\rangle_{\forall j=1,2,3;k_{j},l_{j}=1,2}$.
One can also find the following relation between this construction and $3$-tangle $\tau$ and hyperdeterminat $\mathcal{U}$ which is given by
\begin{eqnarray}\label{HH}
\tau/4=\mathcal{U}&=& d_{1}-2d_{2}+4d_{4}
\end{eqnarray}where
$d_{1}=\alpha^{2}_{000}\alpha^{2}_{111}
+\alpha^{2}_{001}\alpha^{2}_{110}+\alpha^{2}_{010}\alpha^{2}_{101}+
\alpha^{2}_{100}\alpha^{2}_{011}$,
$d_{2}=\alpha_{000}\alpha_{001}\alpha_{110}\alpha_{111}
+\alpha_{000}\alpha_{010}\alpha_{101}\alpha_{111}+
\alpha_{000}\alpha_{100}\alpha_{011}\alpha_{111}
+\alpha_{001}\alpha_{010}\alpha_{101}\alpha_{110}+
\alpha_{001}\alpha_{100}\alpha_{011}\alpha_{110}
+\alpha_{010}\alpha_{100}\alpha_{100}\alpha_{101}$, and
$d_{4}=\alpha_{000}\alpha_{110}\alpha_{101}\alpha_{011}
+\alpha_{111}\alpha_{100}\alpha_{010}\alpha_{001}$
which is a good measure of entanglement for tripartite system. In this expression $d_{1}$ are diagonal lines in the three cube (which we have shown to be the toric variety $\mathcal{X}_{P}$), $d_{2}$ are the diagonal planes, and $d_{4}$ is a tetrahedron.

In summary, we have discussed the construction of affine toric varieties based on polytopes and fan. Then by embedding these combinatorial objects we were able to define the projective toric varieties which are the space of separable state of a quantum system. We have also shown the relation between toric variety construction of three-qubit state and three tangle and hyperdeterminant. Our result can also be of interest in the field of quantum computation, e.g., for the construction of quantum register of a quantum computer.

\end{document}